\documentclass[dvips]{article}
\usepackage{icrctc07}

\title{
Test of hadronic interaction models with data from the Pierre Auger Observatory}

\shorttitle{Test of hadronic interaction models with Auger data}

\authors{Ralph Engel$^1$, for the Pierre Auger Collaboration$^2$}

\shortauthors{R. Engel et al.}

\afiliations{$^1$ Forschungszentrum Karlsruhe, P.O. Box 3640, 76021 Karlsruhe, Germany\\
$^2$ Observatorio Pierre Auger, Av. San Mart\'in Norte 304,
(5613) Malarg\"ue, Mendoza, Argentina}

\email{Ralph.Engel@ik.fzk.de}

\abstract{ The Pierre Auger Observatory allows the measurement of both
  longitudinal profiles and lateral particle distributions of
  high-energy showers. The former trace the overall shower
  development, mainly of the electromagnetic component close to the
  core where the latter reflect the particle densities in the tail of
  the shower far away from the core and are sensitive to both the
  muonic and electromagnetic components. Combining the two
  complementary measurements, predictions of air shower simulations
  are tested. In particular the muon component of the tank signals,
  which is sensitive to hadronic interactions at high energy, is
  studied with several independent methods. Implications for the
  simulation of hadronic interactions at ultra-high energy are
  discussed.  }

\begin{document}
\maketitle

\section{Introduction}

During the last decade, air shower simulation codes have reached
high enough quality that there is good overall agreement between the
predicted and experimentally observed shower characteristics. The
largest remaining source of uncertainty of shower predictions
stems from our limited knowledge of hadronic interactions at high
energy. Hadronic multiparticle production has to be simulated at
energies exceeding by far those accessible at terrestrial accelerators
and in regions of phase space not covered in collider
experiments. Therefore it is not surprising that predictions for the
number of muons or other observables, which are directly related to
hadron production in showers, depend strongly on the adopted hadronic
interaction models \cite{Knapp:2002vs}.

In this work we will employ universality features of the longitudinal
profile of the electromagnetic shower component to combine
fluorescence detector and surface array measurements of the Pierre
Auger Observatory. Using the measured depth of shower maximum, $X_{\rm
  max}$, the muon density at ground is inferred without
assumptions regarding the primary cosmic ray composition. This allows a
direct test of the predictions of hadronic interaction models.

\section{Parameterisation of surface detector signal using universality}

Universality features of the longitudinal profile of showers have been
studied by several authors \cite{BilloirCORSIKASchool05}. Here we exploit
shower universality features to predict the surface detector signal
expected for Auger Cherenkov tanks due to the electromagnetic and
muonic shower components at 1000\,m from the shower core. In the
following only a brief introduction to the method of parameterising the
muonic and electromagnetic tank signals is given. A detailed
description is given in \cite{Schmidt07aa}.

A library of proton and iron showers covering the energy range from
$10^{17}$ to $10^{20}$\,eV and zenith angles between $0^\circ$ and
$70^\circ$ was generated with CORSIKA 6.5 \cite{Heck98a} and the
hadronic interaction models QGSJET II.03 \cite{Ostapchenko:2005nj} and
FLUKA \cite{Fasso01a}. For comparison, a smaller set of showers was
simulated with the combinations QGSJET II.03/GHEISHA
\cite{Fesefeldt85a} and SIBYLL 2.1/FLUKA
\cite{Fletcher:1994bd,Engel99a}. Seasonal models of the Malargue
molecular atmosphere were used \cite{Keilhauer:2004jr}. The detector 
response is calculated using look-up tables derived from a detailed 
GEANT4 simulation \cite{Auger-ICRC07-Henry-Yvon}.

Within the library of showers, the predicted surface detector signal
for the electromagnetic component of a shower at the lateral distance of 1000\,m 
is found to depend mainly on the energy
and the distance between the shower maximum and the ground (distance
to ground, $DG= X_{\rm ground} - X_{\rm max}$). Here the signal of
electromagnetic shower component is defined as that of all shower
particles except muons and decay products of muons.  The signal at
1000\,m depends only slightly on the mass of the primary particle ($13$\% difference
between proton and iron primaries) and the applied interaction model
($\sim 5$\%). The functional form, however, is universal. The
situation is similar for the expected tank signal due to muons and
their decay products. In this case the shower-to-shower fluctuations
are larger and the difference between proton and iron showers amounts
to $40$\%.

\begin{figure}[thb!]
\centerline{
\includegraphics[width=0.4\textwidth,angle=0,clip]{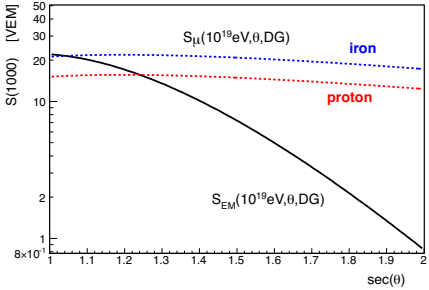}
}
\caption{Electromagnetic and muon contributions to the detector signal
  as a function of zenith angle. Results of QGSJET II/FLUKA
  simulations are shown for $10^{19}$\,eV
  showers.  \label{fig:S1000-signal-contributions}
}
\end{figure}

After accounting for geometrical effects such as the projected tank
surface area, the proton-iron averaged electromagnetic shower signal
is parameterised in dependence on the energy $E$, distance to shower
maximum $DG$, and zenith angle $\theta$. The difference between proton
and iron shower profiles is included in the calculation of the
systematic uncertainties later. Similarly the universal shape of muon
signal profile is parameterized simultaneously for all model primaries, 
taking the overall normalisation from
proton showers simulated with QGSJET II/FLUKA. The expected detector
signal at 1000\,m can then be written as
\begin{eqnarray}
S_{\rm MC}(E,\theta, X_{\rm max}) &=& S_{\rm em}(E,\theta,DG) 
\nonumber\\
& & \hspace*{-2cm}
+ N_\mu^{\rm rel} S_\mu^{\rm QGSII,p} (10^{19}\,{\rm eV}, \theta, DG), 
\label{eq:S1000}
\end{eqnarray}
where $N_\mu^{\rm rel}$ is the number of muons relative to that of
QGSJET proton showers at $10^{19}$\,eV and $S_\mu^{\rm QGSII,p}$ is
the muon signal predicted by QGSJET II for
proton primaries. The relative importance of the electromagnetic and
muonic detector signal contributions at different angles is shown in
Fig.~\ref{fig:S1000-signal-contributions}.

\section{Constant-intensity-cut method}

Within the current statistics, the arrival direction distribution of
high-energy cosmic rays is found to be isotropic, allowing us to apply
the constant intensity cut method to determine the muon signal
contribution. Dividing the surface detector data into equal exposure
bins, the number of showers with $S(1000)$ greater than than a given
threshold should be the same for each bin
\begin{equation}
\left. \frac{dN_{\rm ev}}{d\sin^2 \theta}\right|_{S(1000) > S_{\rm MC}(E,\theta,\langle X_{\rm max} \rangle, 
N_\mu^{\rm rel})} = \rm const.
\label{eq:CIC-muons}
\end{equation} 
Using the independently measured mean depth of shower maximum $\langle
X_{\rm max} \rangle$ \cite{Auger-ICRC07-Unger} the only remaining free
parameter in Eq.~(\ref{eq:CIC-muons}) is the relative number of muons
$N_\mu^{\rm rel}$. For a given energy $E$, $N_\mu^{\rm rel}$ is
adjusted to obtain a flat distribution of events in $\sin^2\theta$.

\begin{figure}[thb!]
\centerline{
\includegraphics[width=0.45\textwidth,angle=0,clip]{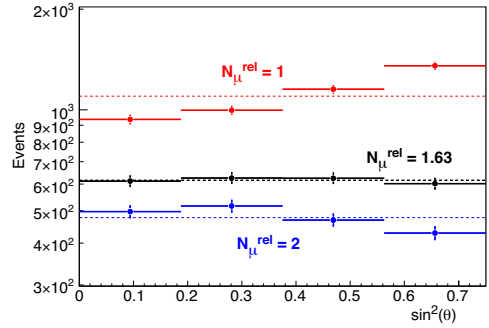}
}
\caption{Sensitivity of the constant-intensity-cut method
  to the muon number for $E=10^{19}$\,eV.
\label{fig:cic-results}
}
\end{figure}

The sensitivity of this method to the muon number parameter in
Eq.~(\ref{eq:S1000}) is illustrated in Fig.~\ref{fig:cic-results}. The
best description of the data above $10^{19}$\,eV requires $N_\mu^{\rm
  rel}= 1.63$. However, this result was obtained by using the measured
mean depth of shower maximum \cite{Auger-ICRC07-Unger} in
Eq.~(\ref{eq:S1000}). Shower-to-shower fluctuations in $X_{\rm max}$
and the reconstruction resolution cannot be neglected and have been
estimated with a Monte Carlo simulation. Accounting for fluctuations
and reconstruction effects, the relative number of muons at
$10^{19}$\,eV is found to be $1.45 \pm 0.11 ({\rm stat}) {}^{+0.11}_{-0.09} ({\rm sys})$.

Knowing the muon number and the measured mean depth of shower maximum,
the signal size at $\theta=38^\circ$ can be calculated
\begin{equation}
S_{38}(10^{19} {\rm eV}) = 37.5 \pm 1.7 ({\rm stat}) {}^{+2.1}_{-2.3} ({\rm sys}) \ {\rm VEM}.
\label{eq:energy-scale}
\end{equation}
This value of $S_{38}$ is a measure of the energy scale of the surface
detector which is independent of the fluorescence detector. It is
within the systematic uncertainties of the energy determination from
fluorescence detector measurements, including the uncertainty of the
fluorescence yield \cite{Auger-ICRC07-Roth}. It corresponds to
assigning showers a $\sim 30$\% higher energy than done in the
fluorescence detector-based Auger shower reconstruction ($E = 1.3
E_{\rm FD}$).

\section{Hybrid event and inclined shower analysis}

Hybrid events that trigger the surface detector array and the
fluorescence telescopes separately are ideally suited to study the
correlation between the depth of shower maximum and the muon density
at 1000\,m. However, the number of events collected so far is much
smaller. For each individual event the reconstructed fluorescence
energy and depth of maximum are available and the expected $S(1000)$
due to the electromagnetic component can be calculated directly. The
difference in the observed signal is attributed to the muon shower
component and compared to the predicted muon signal.

\begin{figure}[thb!]
\centerline{
\includegraphics[width=0.48\textwidth,angle=0,clip]{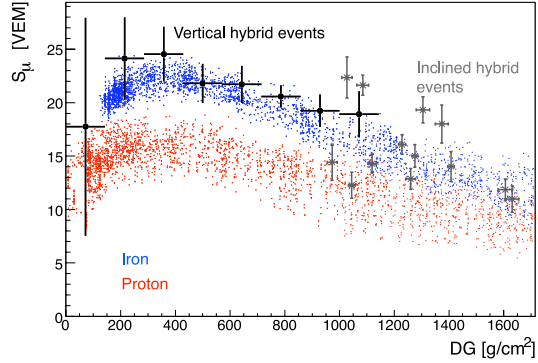}
}
\caption{Reconstructed and predicted muon tank signal contribution
 in dependence on the distance to ground for vertical and inclined hybrid events.
  The muon profiles expected from
  QGSJET II simulations are indicated by the red (proton showers) and blue
  (iron showers) points for the energy scale $E= 1.3 E_{\rm FD}$.
\label{fig:hybrid-muons}
}
\end{figure}

For this study, high-quality hybrid events were selected for which the
shower maximum was in the field of view of a telescope, $\theta <
60^\circ$, and the Mie scattering length was measured. Furthermore the
distance between the telescope and the shower axis was required to be
larger than 10\,km and the Cherenkov light fraction was limited to
less than 50\%. The surface detector event had to satisfy the T5
selection cuts which are also applied in \cite{Auger-ICRC07-Roth}.

In Fig.~\ref{fig:hybrid-muons}, we show the muon signal derived from
these hybrid events as function of distance to ground. The
relative number of muons at $10^{19}$\,eV is found to be
\begin{eqnarray}
\left. N_\mu^{\rm rel}\right|_{E= 1.3 E_{\rm FD}} &=& 1.53 \pm 0.05
\nonumber\\
\left. N_\mu^{\rm rel}\right|_{E= E_{\rm FD}} &=& 1.97 \pm 0.06,
\end{eqnarray}
consistent with the analysis above. 

A similar study has been performed for inclined hybrid events
($60^\circ < \theta < 70^\circ$).  Within the limited statistics, good
agreement between muon numbers of the inclined and the vertical data
sets is found, see Fig.~\ref{fig:hybrid-muons}.

\begin{figure}[thb!]
\centerline{
\includegraphics[width=0.48\textwidth,angle=0,clip]{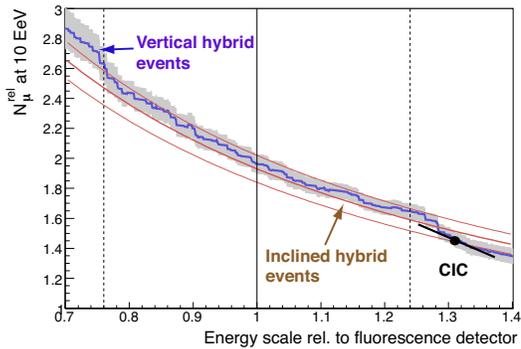}
}
\caption{Comparison of the results on the relative muon multiplicity
  at $10^{19}$\,eV from different methods.  
\label{fig:comparison}
}
\end{figure}

In Fig.~\ref{fig:comparison} we compare the results of the different
methods applied for inferring the muon density at 1000\,m from the
shower core. The relative number of muons is shown as function of the
adopted energy scale with respect to the Auger fluorescence detector
energy reconstruction. Only the constant-intensity-cut method is
independent of the energy scale of the fluorescence detector. Very
good agreement between the presented methods is found.

\section{Discussion}

Assuming universality of the electromagnetic shower component at depths
larger than $X_{\rm max}$, we have determined the muon density and the
energy scale with which the data of the Auger Observatory can be
described self-consistently. The number of muons measured in data is
about 1.5 times bigger than that predicted by QGSJET II for proton
showers. Consistent results were obtained with several analysis
methods.

The QGSJET II and SIBYLL 2.1 predictions for iron showers correspond
to relative muon numbers of 1.39 and 1.27, respectively. Therefore,
interpreted in terms of QGSJET II or SIBYLL 2.1, the derived muon
density would correspond to a primary cosmic ray composition heavier than
iron, which is clearly at variance with the measured $X_{\rm max}$
values. The discrepancy between air shower data and simulations
reported here is qualitatively similar to the inconsistencies found in
composition analyses of previous detectors, see, for example,
\cite{Abu-Zayyad:1999xa,Dova:2004nq,Watson:2004rg}.

Finally it should be mentioned that the results of this study depend
not only on the predictions of the hadronic interaction models but also on 
the reliability of the model used for calculating the
electromagnetic interactions (EGS4 in this study \cite{egs4}).


\begin{thebibliography}{10}

\bibitem{Knapp:2002vs}
J.~Knapp, D.~Heck, S.~J. Sciutto, M.~T. Dova, and M.~Risse,
Astropart. Phys. 19 (2003) 77--99
 and astro-ph/0206414.

\bibitem{BilloirCORSIKASchool05}
P. Billoir, 
lecture given at CORSIKA School, 
http://www-ik.fzk.de/corsika/corsika-school, 2005;
M.~Giller, A.~Kacperczyk, J.~Malinowski, W.~Tkaczyk, and G.~Wieczorek,
J. Phys. G31 (2005) 947--958;
F.~Nerling, J.~Bl{\"u}mer, R.~Engel, and M.~Risse,
Astropart. Phys. 24 (2006) 421--437
 and astro-ph/0506729;
D.~Gora {\it et~al.},
Astropart. Phys. 24 (2006) 484--494
 and astro-ph/0505371;
A.~S. Chou [Pierre Auger Collab.],
Proc. of 29th ICRC (Pune), India, 3-11 Aug
  2005, p. 319.

\bibitem{Schmidt07aa}
F. Schmidt, M. Ave, L. Cazon, A. Chou, these proceedings \#0752,
2007.

\bibitem{Heck98a}
D.~Heck, J.~Knapp, J.~Capdevielle, G.~Schatz, and T.~Thouw,
Wissenschaftliche Berichte FZKA 6019, Forschungszentrum Karlsruhe,
1998.

\bibitem{Ostapchenko:2005nj}
S.~Ostapchenko,
Phys. Rev. D74 (2006) 014026
 and hep-ph/0505259.

\bibitem{Fasso01a}
A.~Fasso, A.~Ferrari, J.~Ranft, and R.~P. Sala,
in Proc. of Int. Conf. on Advanced Monte Carlo for Radiation Physics, Particle
  Transport Simulation and Applications (MC 2000), Lisbon, Portugal, 23-26 Oct
  2000, A. Kling, F. Barao, M. Nakagawa, L. Tavora, P. Vaz eds.,
  Springer-Verlag Berlin, p. 955,
2001.

\bibitem{Fesefeldt85a}
H.~Fesefeldt,
preprint PITHA-85/02, RWTH Aachen,
1985.

\bibitem{Fletcher:1994bd}
R.~S. Fletcher, T.~K. Gaisser, P.~Lipari, and T.~Stanev,
Phys. Rev. D50 (1994) 5710.

\bibitem{Engel99a}
R.~Engel, T.~K. Gaisser, P.~Lipari, and T.~Stanev,
in Proceedings of the 26th ICRC (Salt Lake City)
  vol.~1, p.~415,
1999.

\bibitem{Keilhauer:2004jr}
B.~Keilhauer, J.~Bl{\"u}mer, R.~Engel, H.~O. Klages, and M.~Risse,
Astropart. Phys. 22 (2004) 249
 and astro-ph/0405048.

\bibitem{Auger-ICRC07-Henry-Yvon}
P.L. Ghia and I.Lhenry-Yvon [Pierre Auger Collab.], these proceedings \#0300,
2007.

\bibitem{Auger-ICRC07-Unger}
M.~Unger  [Pierre Auger Collab.], these proceedings \#0594,
2007.

\bibitem{Auger-ICRC07-Roth}
M.~Roth [Pierre Auger Collab.], these proceedings \#0313,
2007.

\bibitem{Abu-Zayyad:1999xa}
T.~Abu-Zayyad {\it et~al.}  (HiRes-MIA Collab.),
Phys. Rev. Lett. 84 (2000) 4276
 and astro-ph/9911144.

\bibitem{Dova:2004nq}
M.T.~Dova, M.E.~Mancenido, A.G.~Mariazzi, T.P.~McCauley and A.A. Watson,
Astropart. Phys. 21 (2004) 597 
 and astro-ph/0312463.

\bibitem{Watson:2004rg}
A.A.~Watson, Nucl. Phys. Proc. Suppl. 136 (2004) 290 
 and astro-ph/0408110.

\bibitem{egs4}
W.R.~Nelson, Report SLAC-265, Stanford Linear Accelerator Center, 
1985.


\end{thebibliography}


\end{document}